\documentclass[aps,prd,amsmath,amssymb,nofootinbib,notitlepage]{revtex4-1}

\usepackage{comment}
\usepackage{color}
\usepackage{amsmath}
\usepackage{amssymb}
\usepackage{amsthm}
\usepackage{mathrsfs}
\usepackage{graphicx}
\usepackage{marvosym}
\usepackage{amsmath,bm}
\usepackage{array}
\usepackage{latexsym}
\usepackage{enumerate}
\usepackage{slashed}
\usepackage{hyperref}
\usepackage{color}
\usepackage{setspace}
\usepackage[margin=20pt,small]{caption}
 \usepackage[normalem]{ulem}

\begin{document}
\fontsize{12pt}{12pt}\selectfont
\title{Side-Jump Induced Spin-Orbit Interaction of Chiral Fluids from Kinetic Theory}
\author{Di-Lun Yang}

\affiliation{RIKEN Nishina Center, RIKEN, Wako, Saitama 351-0198, Japan}
\affiliation{Yukawa Institute for Theoretical Physics, Kyoto University, Kyoto 606-8502, Japan}
\begin{abstract}
We apply the Wigner-function approach and chiral kinetic theory to investigate the angular momentum and polarization of chiral fluids composed of Weyl fermions with background electric/magnetic fields and vorticity. It is found that the quantum corrections in Wigner functions give rise to nonzero anti-symmetric components in the canonical energy-momentum tensors, which are responsible for the spin-orbit interaction. In global equilibrium, conservation of the canonical angular momentum reveals the cancellation between the orbital component stemming from side jumps with nonzero vorticity and the spin component in the presence of an axial chemical potential. We further analyze the conservation laws near local equilibrium. It turns out that the canonical angular momentum is no longer conserved even in the absence of background fields due to the presence of a local torque coming from the spin-orbit interaction involving temperature/chemical-potential gradients, which is implicitly led by collisions. 

\end{abstract}

\maketitle
\section{Introduction}
Recently, there have been intensive studies upon the transport properties of chiral matter composed of Weyl fermions, which involve parity-odd transport such as the chiral magnetic/vortical effects (CME/CVE) in relation to quantum anomalies \cite{Vilenkin:1979ui, Kharzeev:2007jp,Fukushima:2008xe,Landsteiner:2011cp}. In experiments, relativistic heavy ion collisions (HIC) and Weyl semimetals provide the suitable testing grounds for exploring such anomalous transport \cite{Kharzeev:2015znc,Zyuzin:2012tv}. Particularly, recent observations of the negative magneto-resistance in Weyl semimetals suggest the existence of CME \cite{Li:2014bha}. In HIC, the light quarks in quark gluon plasmas (QGP) could be approximated as massless fermions at finite temperature. Despite further interactions with gluons, these quarks move collectively and form a chiral fluid. Such a fluid-like scenario following the charge and energy-momentum conservations in HIC could be rather different from the case in Weyl semimetals. Nevertheless, at high-temperature regime, the fluid-like behaviors of Weyl semimetals have been observed in a recent experiment \cite{Chiral_fluids_WP2}. It is thus intriguing and imperative to further investigate the anomalous transport of chiral fluids. In theory, there exist a variety of approaches to analyze anomalous transport of Weyl fermions including field-theory calculations based on Kubo formula \cite{Fukushima:2008xe,Kharzeev:2009pj,Landsteiner:2011cp}, kinetic theory \cite{Gao:2012ix, Son:2012wh, Stephanov:2012ki, Son:2012zy,Chen:2012ca,Manuel:2013zaa,Chen:2014cla,Chen:2015gta,Hidaka:2016yjf,Hidaka:2017auj,Hidaka:2018ekt,Carignano:2018gqt}, 
relativistic hydrodynamics \cite{Son:2009tf,Neiman:2010zi,Sadofyev:2010pr,Kharzeev:2011ds},
lattice simulations \cite{Abramczyk:2009gb,Buividovich:2009wi,Buividovich:2009zzb,Buividovich:2010tn,Yamamoto:2011gk,Mueller:2016ven,Mace:2016shq}, and
gauge/gravity duality \cite{Erdmenger2009,Torabian2009a,Banerjee2011,Landsteiner2011}. In addition, the anomalous transport induced by rigid-body rotation has been investigated in some theoretical studies \cite{Ambrus:2015lfr,Huang:2017pqe}. Moreover, the recent studies of non-equilibrium anomalous transport upon chiral fluids have incorporated interactions based on the chiral kinetic theory (CKT) \cite{Chen:2015gta,Hidaka:2016yjf,Hidaka:2017auj,Hidaka:2018ekt,Gorbar:2017toh}.    

On the other hand, the observations of global polarization for $\Lambda$ hyperons in HIC \cite{STAR:2017ckg,Adam:2018ivw} have triggered increasing studies upon the spin-polarization formation and angular momenta of relativistic fluids. In fact,
the studies of spin polarization led by global rotation can be traced back to the Barnett effect \cite{PhysRev.6.239} and Einstein–de Haas effect \cite{Einstein_Haas}. In the context pertinent to HIC, a variety of theoretical models were proposed to address the relevant issues such as the microscopic spin-orbital coupling model \cite{Liang:2004ph,Gao:2007bc}, the statistical-hydrodynamic model \cite{Becattini:2009wh,Becattini2013,Becattini2013a,Hayata:2015lga,Becattini:2016gvu,Becattini:2017gcx}, and
the kinetic-theory approach with Wigner functions \cite{Gao:2012ix,Fang:2016vpj,Pang:2016igs}. Also see Ref.\cite{Wang:2017jpl} for a review of some aforementioned approaches. More recently, to understand the spacetime evolution of local polarization and vorticity, the relativistic hydrodynamics with spin-$1/2$ particles has been introduced in Refs.\cite{Florkowski:2017ruc,Florkowski:2017dyn,Florkowski:2018myy,Florkowski:2018ahw}. Nevertheless, the authors therein just focus on massive fermions. There were also related studies for polarized relativistic fluids through an effective-field-theory approach \cite{Montenegro:2017rbu,Montenegro:2017lvf}. For Weyl fermions, the local polarization has been investigated via the Wigner functions in Refs.\cite{Gao:2012ix,Sun:2017xhx}, while the orbital angular momentum and collisions have not been incorporated in the previous studies. Although the polarization density characterized by a Pauli-Lubanski pseudo vector is independent of the orbital contributions in terms of the Wigner-function construction \cite{Becattini2013a,Fang:2016vpj}, the orbital part in fact encodes the angular-momentum transfer between the fluid and internal degrees of freedom such as the spin of quasi-particles. 


In this paper, we employ the CKT in Wigner-function formalism to analyze the interplay between spin and orbital angular momentum near local equilibrium in chiral fluids, which may shed some light upon the spacetime evolution of polarization for Weyl fermions and the role of interactions therein. The paper is organized as the following : In Sec.\ref{WF_section}, we first review the construction of angular-momentum (AM) tensors through the canonical and Belinfante energy-momentum (EM) tensors for spin-$1/2$ fermions with background electric/magnetic fields and further write down the corresponding phase-space distributions via Wigner functions. In Sec.\ref{AM_section}, we then implement the Wigner functions and CKT up to $\mathcal{O}(\hbar)$ to evaluate the AM-tensor density and analyze the angular-momentum conservation with the interplay between spin and orbital components for chiral fluids in global and (near-)local equilibrium. In Sec.\ref{conclude_section}, we finally make short conclusions and outlook.
      
\section{Angular Momenta from Wigner Functions}\label{WF_section}
\subsection{Angular-Momentum Tensors for Fermions}
For simplicity, we will focus on only the dynamics of fermions under background electric/magnetic fields. 
Considering the quantum electrodynamics (QED) Lagrangian with background gauge fields,
\begin{eqnarray}
\mathcal{L}=\bar{\psi}\left(\frac{i\hbar}{2}\gamma^{\mu}\overleftrightarrow{D}_{\mu}-m\right)\psi,
\end{eqnarray} 
where $\overleftrightarrow{D}_{\mu}=\overrightarrow{D}_{\mu}-\overleftarrow{D}^{\dagger}_{\mu}$ and $D_{\mu}=\partial_{\mu}+ieA_{\mu}/\hbar$ denotes the covariant derivative. 
Based on the Noether's theorem and equations of motion, we obtain the canonical EM tensor,
\begin{eqnarray}\nonumber
&&
\bar{T}^{\mu\nu}=T^{\mu\nu}+T^{\mu\nu}_A,\quad
T^{\mu\nu}=\frac{i\hbar}{4}\bar{\psi}\gamma^{\{\mu}\overleftrightarrow{D}^{\nu\}}\psi,
\quad
T^{\mu\nu}_A=\frac{i\hbar}{4}\bar{\psi}\gamma^{[\mu}\overleftrightarrow{D}^{\nu]}\psi,
\end{eqnarray}
and the canonical AM tensor or the so-called mass-energy-momentum tensor or the generalized angular momentum (see e.g. \cite{Becattini:2011ev,Becattini:2012pp,Leader:2013jra}),
\begin{eqnarray}
&&M_C^{\lambda\mu\nu}=M^{\lambda\mu\nu}_{S}+M^{\lambda\mu\nu}_{O},
\\\nonumber
&&M^{\lambda\mu\nu}_{S}=\frac{\hbar}{2}\bar{\psi}\{\gamma^{\lambda},\Sigma^{\mu\nu}\}\psi=-\frac{\hbar}{2}\epsilon^{\lambda\mu\nu\rho}\bar{\psi}\gamma_5\gamma_{\rho}\psi,
\\\nonumber
&&M^{\lambda\mu\nu}_{O}=\frac{i\hbar}{2}\bar{\psi}\gamma^{\lambda}\Big(x^{\mu}\overleftrightarrow{D}^{\nu}
-x^{\nu}\overleftrightarrow{D}^{\mu}\Big)\psi
=x^{\mu}T^{\lambda\nu}-x^{\nu}T^{\lambda\mu}+x^{\mu}T_A^{\lambda\nu}-x^{\nu}T_A^{\lambda\mu}
,
\end{eqnarray}
where $A^{\{\mu}B^{\nu\}}=A^{\mu}B^{\nu}+A^{\nu}B^{\mu}$, $A^{[\mu}B^{\nu]}=A^{\mu}B^{\nu}-A^{\nu}B^{\mu}$, $\Sigma^{\mu\nu}=\frac{i}{4}[\gamma^{\mu},\gamma^{\nu}]$, and $\overleftrightarrow{D}^{\nu}$ only acts on $\psi$ and $\bar{\psi}$. For the canonical EM tensor $\bar{T}^{\mu\nu}$, we decompose it into a symmetric EM tensor $T^{\mu\nu}$ and an anti-symmetric one $T^{\mu\nu}_A$, where $T^{\mu\nu}$ is also known as the Belinfante EM tensor. For the canonical AM tensor, we can also separate the contributions from the spin and orbital angular momentum. Here $M^{\lambda\mu\nu}_{S/O}$ represent the spin/orbital AM tensors. Such a decomposition for the canonical AM tensor is widely utilized in the study of nucleon spins in deep inelastic scattering (DIS) \cite{Jaffe:1989jz,Ji:1996ek} (see Ref.\cite{Leader:2013jra} for a comprehensive review). The canonical AM tensor can be related to the Belinfante AM tensor constructed by only the symmetric EM tensor \cite{Becattini:2011ev,Becattini:2012pp,Leader:2013jra},
\begin{eqnarray}
M^{\lambda\mu\nu}_{B}=x^{\mu}T^{\lambda\nu}-x^{\nu}T^{\lambda\mu}=\frac{i}{4}\bar{\psi}\Big(x^{\mu}\gamma^{\{\lambda}\overleftrightarrow{D}^{\nu\}}
-x^{\nu}\gamma^{\{\lambda}\overleftrightarrow{D}^{\mu\}}\Big)\psi,
\end{eqnarray}
through equations of motion and a total-derivative terms,
\begin{eqnarray}
M_{C}^{\lambda\mu\nu}=M^{\lambda\mu\nu}_{B}+\partial_{\beta}V^{[\beta\lambda][\mu\nu]},
\end{eqnarray}
where $\partial_{\beta}V^{[\beta\lambda][\mu\nu]}$ corresponds to a superpotential antisymmetric in $\lambda,\beta$ and $\mu,\nu$ \footnote{In Ref.\cite{Leader:2013jra}, it is shown $V^{[\beta\lambda][\mu\nu]}=X^{\mu}G^{\beta\lambda\nu}-X^{\nu}G^{\beta\lambda\mu}$, where $G^{\beta\lambda\nu}=\frac{1}{2}\big(M^{\beta\lambda\nu}_S+M^{\nu\beta\lambda}_S+M^{\nu\lambda\beta}_S\big)$. One can thus write down the exact relation between canonical and Belifante AM tensors in phase space in terms of Wigner functions. Furthermore, given Eq.(\ref{MEM_density}) as will be derived shortly, we find $M^{\lambda\mu\nu}_{B}(X)=M^{\lambda\mu\nu}_{C}(X)+\frac{1}{4}\epsilon^{\beta\lambda[\mu\alpha}\partial_{\beta}\big(X^{\nu]}J_{5\beta}(X)\big)$. However, this relation will be further modified when collisions are involved.}. In the absence of background fields, both $M_{C}^{\lambda\mu\nu}$ and $M^{\lambda\mu\nu}_{B}$ are conserved,
\begin{eqnarray}
\partial_{\lambda}M_{C}^{\lambda\mu\nu}=\partial_{\lambda}M^{\lambda\mu\nu}_{B}=0,
\end{eqnarray}
based on the conservation of the symmetric EM tensor. The conservation of $M_{C}^{\lambda\mu\nu}$also implies $\partial_{\lambda}M^{\lambda\mu\nu}_{\text{spin}}=-2T^{\mu\nu}_A$. Accordingly, the anti-symmetric component of $\bar{T}^{\mu\nu}$ serves as a source or sink for spin currents.  
When having background fields peculiarly an electric field or in the case for local equilibrium of relativistic fluids, the conservation laws turn out to be more involved due to collisions. We will further discuss such a case in the later section. 

We may now construct the quantum expectation values for $M^{\lambda\mu\nu}_{C/B}(q,X)$ in phase space via the Wigner-function formalism. Wigner functions are defined as the Wigner transformation of lesser/greater propagators,
\begin{eqnarray}
\grave{S}^{<(>)}(q,X)\equiv\int d^4Ye^{\frac{iq\cdot Y}{\hbar}}S^{<(>)}(x,y),
\end{eqnarray}
where $S^<(x,y)=\langle\psi^{\dagger}(y)\psi(x)\rangle$ and $S^>(x,y)=\langle\psi(x)\psi^{\dagger}(y)\rangle$ as the expectation values of fermionic correlators with $Y=x-y$ and $X=(x+y)/2$. Here the gauge link is implicitly embedded to keep gauge invariance and hence $q_{\mu}$ denotes the kinetic momentum.
For convenience, we will work in the Weyl basis $\psi^{\dagger}=(\psi^{\dagger}_L, \psi^{\dagger}_R)$, which gives
\begin{eqnarray}\nonumber\label{Weyl_rep}
&&T^{\mu\nu}=\frac{i\hbar}{4}\Big(\psi^{\dagger}_R\sigma^{\{\mu}\overleftrightarrow{D}^{\nu\}}\psi_R+\psi^{\dagger}_L\bar{\sigma}^{\{\mu}\overleftrightarrow{D}^{\nu\}}\psi_L\Big),
\quad
T_A^{\mu\nu}=\frac{i\hbar}{4}\Big(\psi^{\dagger}_R\sigma^{[\mu}\overleftrightarrow{D}^{\nu]}\psi_R+\psi^{\dagger}_L\bar{\sigma}^{[\mu}\overleftrightarrow{D}^{\nu]}\psi_L\Big),
\\
&&M^{\lambda\mu\nu}_{S}=-\frac{\hbar}{2}\epsilon^{\lambda\mu\nu\rho}\Big(\psi^{\dagger}_R\sigma_{\rho}\psi_R-\psi^{\dagger}_L\bar{\sigma}_{\rho}\psi_L\Big),
\quad
M^{\lambda\mu\nu}_{O}=x^{\mu}T^{\lambda\nu}-x^{\nu}T^{\lambda\mu}
+x^{\mu}T_A^{\lambda\nu}-x^{\nu}T_A^{\lambda\mu}.
\end{eqnarray}
Based on (\ref{Weyl_rep}), we can now construct the expectation values of EM and AM tensors for Weyl fermions \footnote{For the computation of $\langle M^{\lambda\mu\nu}_{O}\rangle$, we treat $x^{\mu}$ in $M^{\lambda\mu\nu}_{O}$ as the position-space operator, which then acts on both $\psi(x)$ and $\psi^{\dagger}(y)$.},
\begin{eqnarray}\nonumber\label{expect_c}
\langle T^{\mu\nu}\rangle&=&\frac{i\hbar}{4}\text{tr}\Big(\sigma^{\{\mu}\big(D^{\nu\}}_x-D^{\dagger\nu\}}_y\big)S_R^{<}(x,y)
+\bar{\sigma}^{\{\mu}\big(D^{\nu\}}_x-D^{\dagger\nu\}}_y\big)S_L^{<}(x,y)\Big),
\\
\langle T_A^{\mu\nu}\rangle&=&\frac{i\hbar}{4}\text{tr}\Big(\sigma^{[\mu}\big(D^{\nu]}_x-D^{\dagger\nu]}_y\big)S_R^{<}(x,y)
+\bar{\sigma}^{[\mu}\big(D^{\nu]}_x-D^{\dagger\nu]}_y\big)S_L^{<}(x,y)\Big),
\end{eqnarray}
and
\begin{eqnarray}\nonumber
\langle M^{\lambda\mu\nu}_{S}\rangle&=&-\frac{\hbar}{2}\epsilon^{\lambda\mu\nu\rho}\text{tr}\Big(\sigma_{\rho}S^<_R(x,y)-\bar{\sigma}_{\rho}S^<_L(x,y)\big)\Big),
\\
\langle M^{\lambda\mu\nu}_{O}\rangle&=&
\frac{i\hbar}{2}\text{tr}\Big(\big(x^{\mu}D^{\nu}_x-x^{\nu}D^{\mu}_x
-y^{\mu}D^{\dagger\nu}_y+y^{\nu}D^{\dagger\mu}_y\big)\big(S^<_R(x,y)\sigma^{\lambda}+S^<_L(x,y)\bar{\sigma}^{\lambda}\big)\Big).
\end{eqnarray}
One can then perform the Wigner transformation to write down the expectation values in terms of the Wigner functions in phase space. It is useful to exploit the rule of transformation found in \cite{Vasak:1987um} such that
\begin{eqnarray}\label{D_rules}
D_{x\mu}\grave{S}^<(x,y)\rightarrow \Bigg(\frac{\nabla_{\mu}}{2}-i\hbar^{-1}\Pi_{\mu}\Bigg)\grave{S}^<(q,X),\quad D^{\dagger}_{y\mu}\grave{S}^<(x,y)\rightarrow\Bigg(\frac{\nabla_{\mu}}{2}+i\hbar^{-1}\Pi_{\mu}\Bigg)\grave{S}^<(q,X),
\end{eqnarray}
where
\begin{eqnarray}
\nabla_{\mu}=\partial_{\mu}+j_0(\Box)F_{\nu\mu}\partial^{\nu}_{q},\quad \Pi_{\mu}=q_{\mu}+\frac{\hbar}{2}j_1(\Box)F_{\nu\mu}\partial^{\nu}_{q},\quad\Box=\frac{\hbar}{2}\partial_{\rho}\partial^{\rho}_q.
\end{eqnarray} 
We will hereafter use $\partial_{\mu}\equiv\partial/\partial X^{\mu}$ for convenience.
Here $j_0(\Box),j_1(\Box)$ are modified Bessel functions and $\partial_{\rho}$ in $\Box$ only act on the field strength $F_{\nu\mu}$ when having spacetime-dependent background fields.
Making the $\hbar$ expansion, which corresponds to the gradient expansion for $\partial_{\mu}\ll q_{\mu}$, one finds
\begin{eqnarray}\nonumber
\nabla_{\mu}&=&\partial_{\mu}+F_{\nu\mu}\partial^{\nu}_q-\frac{\hbar^2}{24}(\partial_{\rho}\partial^{\rho}_q)^2F_{\nu\mu}\partial^{\nu}_{q}+\mathcal{O}(\hbar^4), 
\\
\Pi_{\mu}&=&q_{\mu}+\frac{\hbar^2}{12}\partial_{\rho}\partial^{\rho}_qF_{\nu\mu}\partial^{\nu}_{q}+\mathcal{O}(\hbar^4).
\end{eqnarray}
Using (\ref{D_rules}) and parameterizing $\grave{S}^<_R=\bar{\sigma}^{\rho}\grave{S}^<_{R\rho}$ and $\grave{S}^<_L=\sigma^{\rho}\grave{S}^<_{L\rho}$, the Wigner transformation of (\ref{expect_c}) yields
\begin{eqnarray}\nonumber\label{expect_WF2}
T^{\mu\nu}(q,X)&=&\Pi^{\{\nu}\grave{S}_V^{<\mu\}}(q,X),
\quad T_A^{\mu\nu}(q,X)=\Pi^{[\nu}\grave{S}_V^{<\mu]}(q,X),
\\\nonumber
M^{\lambda\mu\nu}_{S}(q,X)&=&-\hbar\epsilon^{\lambda\mu\nu\rho}\grave{S}^<_{5\rho}(q,X),
\\
M^{\lambda\mu\nu}_{O}(q,X)&=&X^{\mu}\bar{T}^{\lambda\nu}(q,X)-X^{\nu}\bar{T}^{\lambda\mu}(q,X)
+\hbar\big(\partial^{\mu}_{q} \nabla^{\nu}-\partial^{\nu}_{q} \nabla^{\mu}\big)\grave{S}^{<\lambda}_V(q,X)
,
\end{eqnarray} 
where $\grave{S}^<_{V\mu}=\grave{S}^<_{R\mu}+\grave{S}^<_{L\mu}$, $\grave{S}^<_{5\mu}=\grave{S}^<_{R\mu}-\grave{S}^<_{L\mu}$, and $\bar{T}^{\mu\nu}=T^{\mu\nu}+T_A^{\mu\nu}$. Combining $M^{\lambda\mu\nu}_{S}(q,X)$ and $M^{\lambda\mu\nu}_{O}(q,X)$, one thus obtain the canonical AM tensor in phase space $M^{\lambda\mu\nu}_{C}(q,X)$. After integrating over momentum space, we acquire the canonical AM-tensor density as
\begin{eqnarray}\nonumber\label{MEM_density}
M_{C}^{\lambda\mu\nu}(X)&=&\int\frac{d^4q}{(2\pi)^4}\big(M^{\lambda\mu\nu}_{S}(q,X)+M^{\lambda\mu\nu}_{O}(q,X)\big)
\\
&=&-\frac{\hbar}{2}\epsilon^{\lambda\mu\nu\rho}J_{5\rho}(X)+\Big(X^{\mu}\bar{T}^{\lambda\nu}(X)-X^{\nu}\bar{T}^{\lambda\mu}(X)\Big),
\end{eqnarray}
in which the total-derivative terms in $M^{\lambda\mu\nu}_{O}(q,X)$ do not contribute and the expression is anticipated from its field-theory definition. In addition, by carrying out the same procedure, we also find the Belinfante AM tensor
\begin{eqnarray}
M^{\lambda\mu\nu}_{B}(q,X)
=X^{\mu}T^{\lambda\nu}(q,X)-X^{\nu}T^{\lambda\mu}(q,X)
+\frac{\hbar}{2}\big(\partial^{\mu}_{q} \nabla^{\{\nu}-\partial^{\nu}_{q} \nabla^{\{\mu}\big)\grave{S}^{<\lambda\}}_V(q,X)
,
\end{eqnarray}
and its density
\begin{eqnarray}
M^{\lambda\mu\nu}_{B}(X)=\int\frac{d^4q}{(2\pi)^4}M^{\lambda\mu\nu}_{B}(q,X)
=X^{\mu}T^{\lambda\nu}(X)-X^{\nu}T^{\lambda\mu}(X)
,
\end{eqnarray}
which takes the expected form as well.
\subsection{Angular-Momentum Density and Polarization}
In this subsection, we briefly discuss the particle polarization constructed from AM tensors. 
The usual relativistic angular momentum is defined by integrating $M^{\lambda\mu\nu}_{C/B}(q,X)$ over a spacelike hypersurface in position space, which could be used to define the polarization.
We can thus define the AM density in phase space in terms of a temporal direction characterized by a local timelike vector perpendicular to the hypersurface (In DIS, it is instead taken along a light-cone direction.),
$\bar{n}^{\mu}(X)$, normalized as $\bar{n}^2=1$. The AM densities in phase space are then written as $M^{\mu\nu}_{C/B}(q,X)\equiv\bar{n}_{\lambda}M^{\lambda\mu\nu}_{C/B}(q,X)$.  On the other hand,  
from $M^{\mu\nu}_{C/B}$, we can accordingly introduce the Pauli-Lubanski pseudo vectors,
\begin{eqnarray}
W^{\mu}_{C/B}(q,X)\equiv-\frac{1}{2}\epsilon^{\mu\nu\alpha\beta}\Pi_{\nu}(M_{C/B})_{\alpha\beta}(q,X),
\end{eqnarray}
in which we further promote $q_{\nu}$ to $\Pi_{\nu}$ here as opposed to the usual definition. 
Inversely, we have 
\begin{eqnarray}\label{rel_M_W}
M_{C/B}^{\alpha\beta}(q,X)=\epsilon^{\alpha\beta\mu\nu}\bar{n}_{\mu}\bar{W}_{C/B\nu}(q,X)+\bar{n}^{\alpha}\bar{n}_{\nu}M^{\beta\nu}_{C/B}(q,X)-\bar{n}^{\beta}\bar{n}_{\nu}M^{\alpha\nu}_{C/B}(q,X),
\end{eqnarray}
where $\bar{W}_{C/B}^{\mu}(q,X)\equiv(\Pi\cdot \bar{n})^{-1}W_{C/B}^{\mu}(q,X)$. From (\ref{expect_WF2}), since $\bar{T}^{\mu\nu}=\Pi^{\nu}\grave{S}^{\mu}_V$, it is found 
\begin{eqnarray}\label{WC}
\bar{W}_C^{\mu}(q,X)=\hbar\grave{S}^{<\mu}_5-\frac{\hbar}{2(\bar{n}\cdot\Pi)}\epsilon^{\mu\nu\alpha\beta}\Pi_{\nu}\bar{n}_{\lambda}\partial_{q[\alpha}\nabla_{\beta]}\grave{S}^{<\lambda}_V,
\end{eqnarray}
where we use $(\Pi\cdot\grave{S}^<_5)=0$ from the master equations of Wigner functions \cite{Vasak:1987um,Hidaka:2016yjf}. We also find
\begin{eqnarray}\label{WB}
\bar{W}_B^{\mu}(q,X)=-\epsilon^{\mu\nu\alpha\beta}\Pi_{\nu}X_{\alpha}\grave{S}^{<}_{V\beta}
-\frac{\hbar}{4(\bar{n}\cdot\Pi)}\epsilon^{\mu\nu\alpha\beta}\Pi_{\nu}\bar{n}_{\lambda}\partial_{q[\alpha}\nabla_{\beta]}\grave{S}^{<\lambda}_V.
\end{eqnarray}
The canonical Pauli-Lubanski pseudo vector is usually proposed to define the polarization of particles in HIC \cite{Becattini2013a,Fang:2016vpj}. One may utilize $\bar{W}(q,X)$ to define the polarization vector \footnote{For massive fermions, one may consider the normalization by the mass of fermions instead of $\bar{n}\cdot q$ \cite{Becattini2013a}.}.
By integrating over momentum space in Eq.(\ref{WC}), one may define the polarization density characterized by just the axial-charge current \cite{Fang:2016vpj},
\begin{eqnarray}\label{LC}
L_C^{\mu}(X)\equiv\int\frac{d^4q}{(2\pi)^4}\bar{W}_C^{\mu}(q,X)=\frac{\hbar}{2}J^{\mu}_5(X).
\end{eqnarray} 
In a particular case when the superindices $\alpha,\beta$ in $M_{C/B}^{\alpha\beta}(q,X)$ are the spatial directions transverse to $\bar{n}^{\mu}$, one could have consistent definitions for polarization in terms of either $M_{C/B}^{\alpha\beta}(q,X)$ or $W^{\mu}_{C/B}(q,X)$ through the relation,   
\begin{eqnarray}
P^{\alpha}_{(\bar{n})\alpha'}P^{\beta}_{(\bar{n})_{\beta'}}M_{C/B}^{\alpha'\beta'}(q,X)=\epsilon^{\alpha\beta\mu\nu}\bar{n}_{\mu}\bar{W}_{C/B\nu}(q,X)
\end{eqnarray}
where $P^{\alpha}_{(\bar{n})\alpha'}=\eta^{\alpha}_{\alpha'}-\bar{n}^{\alpha}\bar{n}_{\alpha'}$ with $\eta_{\mu\nu}$ being the Minkowski-spacetime metric.
As shown in Eq.(\ref{WC}), the last term starting at $\mathcal{O}(\hbar)$ coming from the orbital angular momentum could also contribute to the spectrum of polarization for both massless and massive fermions, which will be more practical for the phenomenology in HIC. Nonetheless, we will investigate the spectrum of polarization elsewhere and only focus on the study of AM-tensor density in this paper.

\section{Spin-Orbit Interaction of Chiral Fluids}\label{AM_section}
\subsection{Global Equilibrium}
We may now implement the thermal-equilibrium Wigner functions perturbatively derived from Kadanoff-Baym-like (KB-like) equations shown in Refs.\cite{Hidaka:2016yjf,Hidaka:2017auj,Hidaka:2018ekt} to compute the AM-tensor density and analyze angular-momentum conservation of chiral fluids composed of Weyl fermions. For simplicity, we firstly consider the global-equilibrium case with constant temperature and chemical potentials in the presence of only magnetic fields and fluid vorticity. 

The Wigner function for right-handed fermions up to $\mathcal{O}(\hbar)$ was derived in Ref.\cite{Hidaka:2016yjf}, 
\begin{eqnarray}\label{Wigner_f}
\grave{S}_{R}^{<\mu}&=&2\pi\bar{\epsilon}(q\cdot n)\Bigg(\delta(q^2)\Big(q^{\mu}
+\hbar S^{\mu\nu}_{(n)}\mathcal{D}_{\nu}\Big)
+\hbar\epsilon^{\mu\nu\alpha\beta}q_{\nu}F_{\alpha\beta}\frac{\partial\delta(q^2)}{2\partial q^2}
\Bigg)f^{(n)}_q
\end{eqnarray}
where
\begin{eqnarray}\label{S_n_1}
S^{\mu\nu}_{(n)}=\frac{\epsilon^{\mu\nu\alpha\beta}}{2(q\cdot n)}q_{\alpha}n_{\beta} 
\end{eqnarray}
corresponds to the spin tensor depending on a frame vector $n^{\mu}$. Here we denote $\mathcal{D}_{\beta}f^{(n)}_q=\Delta_{\beta}f^{(n)}_q-\mathcal{C}_{\beta}$, where $\Delta_{\mu}=\partial_{\mu}+F_{\nu\mu}\partial^{\nu}_q$,  $\mathcal{C}_{\beta}=\Sigma_{\beta}^<\bar{f}^{(n)}_q-\Sigma_{\beta}^>f^{(n)}_q$ with $\Sigma_{\beta}^{<(>)}$ being lesser/greater self-energies and $f^{(n)}_q$ and $\bar{f}^{(n)}_q=1-f^{(n)}_q$ being the distribution functions of incoming and outgoing particles, respectively. Also, the frame vector $n^{\mu}$ comes from the choice of the spin basis, which is different from $\bar{n}^{\mu}$ for fixing the temporal direction in local spacetime although one can set $n^{\mu}=\bar{n}^{\mu}$ for particular conditions. In addition, $\bar{\epsilon}(q\cdot n)$ represents the sign of $q\cdot n$. The $\mathcal{O}(\hbar)$ terms in (\ref{Wigner_f}) contributes to the leading-order quantum corrections for the anomalous transport, in which the spin-tensor-dependent term dubbed as the side-jump term engenders the magnetization current and partial contribution of the CVE. On the other hand, the last term in (\ref{Wigner_f}) leads to the CME in equilibrium and the modified dispersion relation from the magnetic-moment coupling in CKT. It is found that the global-equilibrium distribution function takes the form \cite{Chen:2015gta,Hidaka:2017auj}
\begin{eqnarray}\label{global_equil_f}
f^{\text{eq}(n)}_q=(e^{g}+1)^{-1},\quad g=\left(\beta q\cdot u- \bar{\mu}_R+\frac{\hbar S^{\mu\nu}_{(n)}}{2}\partial_{\mu}(\beta u_{\nu})\right),
\end{eqnarray} 
where $\beta=1/T$ is the inverse of temperature $T$ , $\bar{\mu}_R=\mu_R/T$ with $\mu_R$ being a charge chemical potential for right-handed fermions, and $u^{\mu}$ represents the fluid velocity. Now, Eq.(\ref{Wigner_f}) and Eq.(\ref{global_equil_f}) result in the global-equilibrium Wigner functions for right-handed fermions \cite{Hidaka:2017auj}, 
\begin{eqnarray}\label{WF_R}
\grave{S}_{R\text{geq}}^{<\mu}&=&2\pi\bar{\epsilon}(q\cdot u)\Bigg[\delta(q^2)\left(q^{\mu}+\frac{\hbar}{2}\big(u^{\mu}(q\cdot\omega)-\omega^{\mu}(q\cdot u)\big)\partial_{q\cdot u}
\right)
+\frac{\hbar\epsilon^{\mu\nu\alpha\beta}F_{\alpha\beta}}{4}\partial_{q\nu}\delta(q^2)
\Bigg]f^{(0)}_q,
\end{eqnarray}
where  $f^{(0)}_q=(e^{\beta(q\cdot u-\mu_R)}+1)^{-1}$ corresponds to the usual Fermi-Dirac distribution function and $\omega^{\mu}$ denotes the fluid vorticity defined as
\begin{eqnarray}
\omega^{\mu}\equiv\frac{1}{2}\epsilon^{\mu\nu\alpha\beta}u_{\nu}(\partial_{\alpha}u_{\beta}).
\end{eqnarray} 
Note that Eq.(\ref{WF_R}) is independent of the choice of $n^{\mu}$. For left-handed fermions, the sign in front of each $\mathcal{O}(\hbar)$ term should flip.

Given Eq.(\ref{WF_R}), we may evaluate the EM/AM tensor densities, which are independent of the temporal direction $\bar{n}^{\mu}$. Carrying out the explicit computations of the symmetric EM tensor and charge current in global equilibrium, for the first-order inviscid hydrodynamics of right-handed fermions up to $\mathcal{O}(\hbar)$, it is found \cite{Hidaka:2017auj}
\begin{eqnarray}\nonumber
T_{R\text{geq}}^{\mu\nu}&=&u^{\mu}u^{\nu}\epsilon_R-p_R\Theta^{\mu\nu}+\Pi_{R\text{non}}^{\mu\nu}=\int\frac{d^4q}{(2\pi)^4}\big(q^{\mu}\grave{S}^{<\nu}_{R\text{geq}}+q^{\nu}\grave{S}^{<\mu}_{R\text{geq}}\big),
\\
J_{R\text{geq}}^{\mu}&=&N_{R}u^{\mu}+v_{R\text{non}}^{\mu}=2\int \frac{d^4q}{(2\pi)^4}\grave{S}_{R\text{geq}}^{<\mu},
\end{eqnarray}
where $\Theta^{\mu\nu}=\eta^{\mu\nu}-u^{\mu}u^{\nu}$ and the non-dissipative quantum corrections in global equilibrium take the form
\begin{eqnarray}
\Pi_{R\text{non}}^{\mu\nu}=\hbar\xi_{\omega R}\big(\omega^{\mu}u^{\nu}+\omega^{\nu}u^{\mu}\big)
+\hbar\xi_{BR}\big(B^{\mu}u^{\nu}+B^{\nu}u^{\mu}\big),
\quad v_{R\text{non}}^{\mu}=\hbar\sigma_{BR} B^{\mu}+\hbar\sigma_{\omega R}\omega^{\mu},
\end{eqnarray}
and the transport coefficients read
\begin{eqnarray}\label{coefficients1}\nonumber
\epsilon_R&=&3p_R=T^4\Big(\frac{7\pi^2}{120}+\frac{\bar{\mu}_R^2}{4}+\frac{\bar{\mu}_R^4}{8\pi^2}\Big),
\quad N_{R}=\frac{T^3}{6}\left(\bar{\mu}_R+\frac{\bar{\mu}_R^3}{\pi^2}\right),
\quad
\sigma_{\omega R}=\frac{T^2}{12}\Bigg(1+\frac{3\bar{\mu}_R^2}{\pi^2}\Bigg)
,
\\
\sigma_{BR}&=&\frac{\mu_R}{4\pi^2},\quad
\xi_{\omega R}=\frac{T^3}{6}\Big(\bar{\mu}_R+\frac{\bar{\mu}_R^3}{\pi^2}\Big)=N_{R},\quad \xi_{BR}=\frac{ T^2}{24}\Big(1+\frac{3\bar{\mu}_{R}^2}{\pi^2}\Big)=\frac{\sigma_{\omega R}}{2}.
\end{eqnarray} 
Note that the electric/magnetic fields in this paper are defined with respect to the fluid velocity,
\begin{eqnarray}\label{EM_def}
u^{\nu}F_{\mu\nu}=E_{\mu},\quad \frac{1}{2}\epsilon^{\mu\nu\alpha\beta}u_{\nu}F_{\alpha\beta}=B^{\mu},\quad F_{\alpha\beta}=-\epsilon_{\mu\nu\alpha\beta}B^{\mu}u^{\nu}+u_{\beta}E_{\alpha}-u_{\alpha}E_{\beta}.
\end{eqnarray}

Nonetheless, unlike the case for massive fermions \cite{Becattini2013a,Florkowski:2017dyn}, in which the quantum corrections only come from distribution functions, the anti-symmetric EM tensor is nonzero for Weyl fermions. By using Eq.(\ref{WF_R}), we find \footnote{In fact, the side-jump term and the delta-function-derivative term both lead to the $\hbar$ corrections with magnetic fields on $T^{\mu\nu}_{A\text{geq}}$, whereas these contributions exactly cancel each other. See Eq.(\ref{TA0i}) for details.}
\begin{eqnarray}
T^{\mu\nu}_{AR\text{geq}}=\int \frac{d^4q}{(2\pi)^4}\big(q^{\nu}\grave{S}^{<\mu}_{R\text{geq}}-q^{\mu}\grave{S}^{<\nu}_{R\text{geq}}\big)=-\frac{\hbar}{2}\xi_{\omega R}(u^{\mu}\omega^{\nu}-u^{\nu}\omega^{\mu}).
\end{eqnarray}
Note that such a non-vanishing $T^{\mu\nu}_{AR\text{geq}}$ originates from the side-jump term. Our finding also agrees with what has been recently found in Ref.\cite{Buzzegoli:2018wpy} by the density operator approach.
Combining the symmetric and anti-symmetric parts, we thus obtain
\begin{eqnarray}
\bar{T}^{\mu\nu}_{R\text{geq}}=u^{\mu}u^{\nu}\epsilon_R-p_R\Theta^{\mu\nu}
+\hbar\xi_{B R}\big(B^{\mu}u^{\nu}+B^{\nu}u^{\mu}\big)
+\frac{\hbar}{2}\xi_{\omega R}\big(3\omega^{\mu}u^{\nu}+\omega^{\nu}u^{\mu}\big)
\end{eqnarray}
for right-handed fermions. 
Now, incorporating also the contributions from left-handed fermions, for which the $\mathcal{O}(\hbar)$ corrections should flip the signs, the vector/axial-charge currents read
\begin{eqnarray}\label{JVA}
J^{\mu}_{V/5\text{geq}}=J^{\mu}_{R\text{geq}}\pm J^{\mu}_{L\text{geq}} =N_{V/5}u^{\mu}+\hbar\sigma_{BV/5}B^{\mu}+\hbar\sigma_{\omega V/5}\omega^{\mu},
\end{eqnarray}
where 
\begin{eqnarray}\nonumber
&&
N_{V}=\frac{\mu_V}{6}\Big(T^2+\frac{3\mu_5^2+\mu_V^2}{4\pi^2}\Big),\quad  N_5=\frac{\mu_5}{6}\Big(T^2+\frac{3\mu_V^2+\mu_5^2}{4\pi^2}\Big),
\\
&&\sigma_{BV/5}=\frac{\mu_{5/V}}{4\pi^2},\quad \sigma_{\omega V}=\frac{\mu_V\mu_5}{4\pi^2},\quad
\sigma_{\omega 5}=\Big(\frac{T^2}{6}+\frac{\mu_V^2+\mu_5^2}{8\pi^2}\Big),
\end{eqnarray}
and $\mu_{V/5}=\mu_R\pm\mu_L$. These charge currents are nothing but the CME, CSE, and CVE. Note that the axial-charge chemical potential $\mu_5$ can be regarded as a spin chemical potential since the spin component of the angular momentum is characterized by the axial-charge current. Furthermore, we find 
\begin{eqnarray}\label{TAV}
	T_{A\text{geq}}^{\mu\nu}=\bar{T}_{AR\text{geq}}^{\mu\nu}+\bar{T}_{AL\text{geq}}^{\mu\nu}
	=-\frac{\hbar}{2}\xi_{\omega V}\big(\omega^{\nu}u^{\mu}-\omega^{\mu}u^{\nu}\big)
	,
\end{eqnarray}
and thus
\begin{eqnarray}\nonumber
\bar{T}_{\text{geq}}^{\mu\nu}&=&\bar{T}_{R\text{geq}}^{\mu\nu}+\bar{T}_{L\text{geq}}^{\mu\nu}
\\
&=&u^{\mu}u^{\nu}\epsilon_V-p_V\Theta^{\mu\nu}
+\hbar\xi_{BV}\big(B^{\mu}u^{\nu}+B^{\nu}u^{\mu}\big)+\frac{\hbar}{2}\xi_{\omega V}\big(3\omega^{\mu}u^{\nu}+\omega^{\nu}u^{\mu}\big)
\end{eqnarray}
where $\epsilon_V=3p_V=\epsilon_R+\epsilon_L$ is parity even while
\begin{eqnarray}
\xi_{B V}=\xi_{B R}-\xi_{B L}=\frac{\mu_V\mu_5}{8\pi^2},
\quad
\xi_{\omega V}=\xi_{\omega R}-\xi_{\omega L}=\frac{\mu_5}{6}\Big(T^2+\frac{1}{4\pi^2}(3\mu_V^2+\mu_5^2)\Big)=N_5
\end{eqnarray}
are parity-odd.
By inserting $J^{\mu}_A$ and $\bar{T}^{\mu\nu}_V$ into Eq.(\ref{MEM_density}), one is able to write down the spin/orbital AM-tensor density, $M^{\lambda\mu\nu}_{S/O}(X)$. Eventually, in global equilibrium, one finds
\begin{eqnarray}\nonumber\label{M_so_Geq}
M^{\lambda\mu\nu}_{S\text{geq}}(X)&=&\frac{\hbar}{2}\epsilon^{\lambda\mu\nu\rho}\big(N_5u_{\rho}
+\hbar\sigma_{B5}B_{\rho}+\hbar\sigma_{\omega 5}\omega_{\rho}
\big),
\\
M^{\lambda\mu\nu}_{O\text{geq}}(X)&=&X^{[\mu}\Bigg[\epsilon_V\Big(u^{\lambda}u^{\nu]}-\frac{\Theta^{\lambda\nu]}}{3}\Big)
+\hbar\xi_{BV}\big(B^{\lambda}u^{\nu]}+B^{\nu]}u^{\lambda}\big)
+\frac{\hbar}{2}N_5\big(3\omega^{\lambda}u^{\nu]}+\omega^{\nu]}u^{\lambda}\big)
\Bigg].
\end{eqnarray}
For $M^{\lambda\mu\nu}_{S\text{geq}}(X)$, the leading-order term simply comes from nonzero axial- charge (spin) density, while the axial-charge currents from CSE/CVE yield sub-leading effects. However, when $\mu_5=0$, the CSE/CVE contribution should dominate over $M^{\lambda\mu\nu}_{S\text{geq}}(X)$ and accordingly over $L^{\mu}_{C\text{geq}}(X)$ in Eq.(\ref{LC}). For $M^{\lambda\mu\nu}_{O\text{geq}}(X)$, the first term in (\ref{M_so_Geq}) corresponds to just classical rotation of fluids, while the $\hbar$ corrections are related to CME/CVE in $T_{\text{geq}}^{\mu\nu}$ and the $\omega$-dependent $T^{\mu\nu}_{A\text{geq}}$, which exist only when $\mu_5\neq 0$.    

In global equilibrium, it is trivially to show the conservation of charge currents and symmetric EM tensor, $\partial_{\mu}J^{\mu}_{V/5\text{geq}}=\partial_{\mu}T^{\mu\nu}_{\text{geq}}=0$. Nevertheless, it is now nontrivial to show the conservation of the canonical angular momentum. Using Eq.(\ref{TAV}), we firstly check $\partial_{\mu}T^{\mu\nu}_{A\text{geq}}=0$ based on $\partial\cdot\omega=0$ and $u\cdot\partial\omega^{\mu}=0$ in global equilibrium. One thus obtain
\begin{eqnarray}\label{cons_AM}
\partial_{\lambda}M_{C\text{geq}}^{\lambda\mu\nu}=-\frac{\hbar}{2}\epsilon^{\lambda\mu\nu\rho}\partial_{\lambda}(J_{5\text{geq}})_{\rho}+2T^{\mu\nu}_{A\text{geq}}.
\end{eqnarray}
By taking $J_{5\text{geq}}^{\rho}=N_5u^{\rho}$ and Eq.(\ref{TAV}), one also derives $\partial_{\lambda}M_{C\text{geq}}^{\lambda\mu\nu}=0$ up to $\mathcal{O}(\hbar)$. It is found that the spin part is not conserved by itself at the leading order, which has to be compensated by the orbital angular momentum particularly from the side-jump contribution. The non-vanishing $T^{\mu\nu}_{A\text{geq}}$ here plays a role for the angular-momentum transfer. Physically, it is understood that the collectively orbiting Weyl fermions, which contribute to the rotation of fluids, change the direction of a net spin for the fluid cell since the spin directions are enslaved by the moving directions based on the conservation of helicity.   
When $\mu_5=0$, from Eq.(\ref{JVA}), one finds
\begin{eqnarray}
\partial_{\lambda}M_{S\text{geq}}^{\lambda\mu\nu}=\frac{\hbar^2}{2}\big(\epsilon^{\kappa\mu\nu\rho}\Theta_{\kappa}^{\mbox{ }\lambda}(\sigma_{B5}\partial_{\lambda}B_{\rho}+\sigma_{\omega 5}\partial_{\lambda}\omega_{\rho})+\sigma_{B5}(B^{\mu}\omega^{\nu}-B^{\nu}\omega^{\mu})\big),
\end{eqnarray} 
where we employ $u\cdot\partial\omega^{\mu}=0$ and $u\cdot\partial B^{\mu}=\epsilon^{\mu\nu\alpha\beta}u_{\nu}B_{\alpha}\omega_{\beta}$ in global equilibrium from Bianchi identities. However, to derive the orbital angular momentum at the same order, we have to apply the Wigner functions up to $\mathcal{O}(\hbar^2)$, which have not been derived so far in literature. It is anticipated that this unknown $T^{\mu\nu}_{A\text{geq}}$ at $\mathcal{O}(\hbar^2)$ should cancel the spin part and preserve the total angular momentum.  

\subsection{(Near-)Local Equilibrium}
In local equilibrium with inhomogeneous temperature and chemical potentials, the interaction between Weyl fermions is involved and the conservation laws do not trivially hold. Nevertheless, these conservation laws can be determined by the kinetic theory given the details of collisions. For Weyl fermions, it is again more convenient to firstly work in the right/left-handed bases and later combined the results from two sectors. For right-handed fermions, it is shown in Ref.\cite{Hidaka:2017auj} that the local-equilibrium distribution function $f^{\text{leq}(u)}_q$ takes the same form as Eq.(\ref{global_equil_f}) in the co-moving frame, $n^{\mu}=u^{\mu}$. The non-dissipative anomalous transport for $J^{\mu}_{V/5}$ and $T^{\mu\nu}$ also remains unchanged in local equilibrium. However, there exist dissipative corrections coming from interactions. To acquire a general feature with manifest interpretation in physics, we may simplify the collisional kernels by employing the relaxation-time approximation (RTA). 
Moreover, we also neglect the interactions between right/left handed fermions for simplicity. In such an approximation near local equilibrium, the CKT for right-handed fermions can be simplified as \cite{Hidaka:2017auj,Hidaka:2018ekt}
\begin{eqnarray}\label{CKT}
\Box(q,X)f^{(u)}_q\approx -\frac{q\cdot u}{\tau_R}\delta f_q,
\end{eqnarray}   
where $\delta f_q=f^{(u)}_q-f^{\text{leq}(u)}_q$ denotes the deviation of the distribution function from local equilibrium and
\begin{eqnarray}
\Box(q,X)&=&\Big[
q\cdot\Delta+\hbar\frac{S_{(u)}^{\mu\nu}E_{\mu}}{(q\cdot u)}\Delta_{\nu}
+\hbar S_{(u)}^{\mu\nu}(\partial_{\mu}F_{\rho \nu})\partial^{\rho}_{q}
+\hbar(\partial_{\mu}S^{\mu\nu}_{(u)})\Delta_{\nu}\Big].
\end{eqnarray}
The constant $\tau_R$ represents the relaxation time charactering the inverse strength of interactions between the Weyl fermions with same chirality. We may write the near-local-equilibrium deviations on the symmetric EM tensor and charge four currents as $\delta T^{\mu\nu}=T^{\mu\nu}-T^{\mu\nu}_{\text{leq}}$ and $\delta J^{\mu}_{V/5}=J^{\mu}_{V/5}-J^{\mu}_{V/5\text{leq}}$. Here $T^{\mu\nu}_{\text{leq}}$ and $J^{\mu}_{V/5\text{leq}}$ have the same expressions as global-equilibrium ones by simply replacing constant thermodynamic parameters therein to the local-equilibrium ones with spatial inhomogeneity. In the following computations, we also apply the gradient expansion and only preserve the 1st-order-derivative ($\mathcal{O}(\partial)$) terms in the non-equilibrium deviations. 

Solving Eq.(\ref{CKT}) for $f^{(u)}_q$ and plugging the solution into Wigner functions, after combining with the contribution from left-handed fermions, the non-equilibrium vector/axial-charge currents take the form,
\begin{eqnarray}\label{J_V5_noneq}
\delta J^{\mu}_{V/5\perp}=\frac{\tau_R}{3}\big(\sigma_{\omega 5/V}\mathcal{E}^{\mu}_{V\perp}+\sigma_{\omega V/5}\mathcal{E}^{\mu}_{5\perp}\big)+N_{V/5}\tau_R\Big(\frac{1}{T}\partial^{\mu}_{\perp}T-u\cdot \partial u^{\mu}_{\perp}\Big)+\mathcal{O}(\hbar)
,
\end{eqnarray}
where $V^{\mu}_{\perp}=\Theta^{\mu}_{\mbox{ }\nu}V^{\nu}$ for an arbitrary (pesudo-)vector $V^{\mu}$ and $\mathcal{E}_{V/5\mu}=E_{\mu}+T\partial_{\mu}\bar{\mu}_{V/5}$. 
Here we drop the higher-derivative terms and quantum corrections since the classical part has already led to the $\mathcal{O}(\hbar)$ contribution in the angular momentum. The axial-charge current induced by electric fields is also dubbed as the chiral electric separation effect (CESE) \cite{Huang:2013iia}, for which the corresponding conductivity has been computed in QED plasmas and weakly coupled QGP and as well in holographic models \cite{Huang:2013iia,Jiang:2014ura,Pu:2014cwa,Pu:2014fva}. In addition, the non-equilibrium deviation upon the symmetric EM tensor contains only viscous corrections up to $\mathcal{O}(\partial)$. More precisely, one finds $\Pi^{\mu\nu}=\Theta^{\mu\alpha}\Theta^{\nu\beta}\delta T_{\alpha\beta}=\zeta(\partial\cdot u)\Theta^{\mu\nu}+\eta_s \pi^{\mu\nu}$, where
$\pi^{\mu\nu}\equiv \Theta^{\mu}_{\rho}\Theta^{\nu}_{\sigma}(\partial^{\rho}u^{\sigma}
+\partial^{\sigma}u^{\rho}-2\eta^{\rho\sigma}\theta/3)/2$ denotes the shear strength and $\eta_s/\zeta$ correspond to shear/bulk viscosities \cite{Romatschke:2009im}. Here $\eta_s/\zeta$ depend on $\tau_R$, while their explicit forms are not important in our study. Although there exist no non-equilibrium $\mathcal{O}(\hbar)$ corrections up to $\mathcal{O}(\partial)$, the $\mathcal{O}(\hbar)$ corrections will set in at $\mathcal{O}(\partial^2)$ including anomalous Hall effects and viscous corrections on CME/CVE \cite{Hidaka:2017auj,Hidaka:2018ekt}. Nonzero $ \Pi^{\mu\nu}$ simply contributes to part of the orbital angular momentum conserved independently and irrelevant to the spin component in hydrodynamics. Note that we omit the computations of  non-equilibrium charge densities and energy-density current since they should vanish according to the matching conditions as discussed later. 

On the other hand, by carrying out an explicit calculation with the local-equilibrium Wigner functions, we obtain the non-vanishing $T^{\mu\nu}_{A}$ depending on also electric fields and temperature/chemical-potential gradients,
\begin{eqnarray}\label{TA_leq}
T_{A}^{\mu\nu}=-\frac{\hbar}{2}\xi_{\omega V}(u^{\mu}\omega^{\nu}-u^{\nu}\omega^{\mu})
-\frac{\hbar\epsilon^{\mu\nu\alpha\beta} u_{\alpha}}{6}\Big(\sigma_{\omega V}T\partial_{\beta}\bar{\mu}_V+\sigma_{\omega 5}T\partial_{\beta}\bar{\mu}_5
+3N_5\frac{\partial_{\beta}T}{T}-\sigma_{\omega V}E_{\beta}\Big),
\end{eqnarray}  
where we utilize
\begin{eqnarray}\label{TAR_leq}
T_{AR/L}^{\mu\nu}=\mp\frac{\hbar}{2}\xi_{\omega R/L}(u^{\mu}\omega^{\nu}-u^{\nu}\omega^{\mu})
\mp\frac{\hbar\epsilon^{\mu\nu\alpha\beta} u_{\alpha}}{2}\Big(\sigma_{\omega R/L}T\partial_{\beta}\bar{\mu}_{R/L}+N_{R/L}\frac{\partial_{\beta}T}{T}-\frac{\sigma_{\omega R/L}\mathcal{E}_{R/L\beta}}{3}\Big)
\end{eqnarray}  
as derived in Eq.(\ref{TAij}).
It is clear to see that $T^{\mu\nu}_A\neq\frac{\hbar}{4}\epsilon^{\lambda\mu\nu\rho}\partial_{\lambda}J_{5\rho}$ near local equilibrium and hence $T^{\mu\nu}_A$ and $M^{\lambda\mu\nu}_C$ are no longer conserved. However, such non-conservation has been foreseen by the KB-like equations as the master equations for Wigner functions and CKT shown in Eq.(\ref{TA_Collisions}), from which we derive
\begin{eqnarray}\label{TAC}
T^{\mu\nu}_{A}=\frac{\hbar}{4}\epsilon^{\mu\nu\alpha\beta}\Big(\partial_{\alpha}J_{5\beta}+2n_{\alpha}\int_q\big((q\cdot n)\mathcal{C}_{5\perp\beta}-n\cdot \mathcal{C}_5q_{\perp\beta}\big)\Big)
=\frac{\hbar}{4}\epsilon^{\mu\nu\alpha\beta}\Big(\partial_{\alpha}J_{5\beta}+\frac{u_{\alpha}\delta J_{5\perp\beta}}{\tau_R}\Big),
\end{eqnarray}
where we take the RTA to acquire the second equality. It turns out that the near-local-equilibrium corrections on $T^{\mu\nu}_A$ implicitly depend on collisions even though they can be directly derived from local-equilibrium Wigner functions. By using Eq.(\ref{J_V5_noneq}), one can check that Eq.(\ref{TAC}) agrees with Eq.(\ref{TAR_leq}) up to $\mathcal{O}(\hbar\partial)$. 

Moreover, by utilizing the results in Refs.\cite{Hidaka:2017auj,Hidaka:2018ekt}, in the RTA, the CKT yields the following conservation laws or the so-called matching conditions,
\begin{eqnarray}\label{cons_J_T}
\partial_{\mu}J^{\mu}_V=-\frac{u_{\mu}\delta J_V^{\mu}}{\tau_R},\quad
\partial_{\mu}J^{\mu}_5=-\frac{\hbar E\cdot B}{2\pi^2}-\frac{u_{\mu}\delta J_5^{\mu}}{\tau_R},
\quad\partial_{\mu}T^{\mu\nu}=F^{\nu\rho}J_{V\rho}-\frac{u_{\mu}\delta T^{\mu\nu}}{\tau_R}.
\end{eqnarray}
When a system respects the charge and energy-momentum conservation, we should impose $u_{\mu}\delta J^{\mu}_{V/5}=0$ and $u_{\mu}\delta T^{\mu\nu}=0$, which allows us to define the local-equilibrium temperature, chemical potentials, and fluid velocity.
Then, solving the six conservation equations in Eq.(\ref{cons_J_T}) with constitutive equations in anomalous hydrodynamics yields the hydrodynamic equations of motion (EOM), which give rise to the temporal derivatives with respect to the fluid velocity on six thermodynamic parameters, $u\cdot\partial T$, $u\cdot\partial \bar{\mu}_{V/5}$, and $u\cdot\partial u^{\mu}$. Note that the temporal component of $u^{\mu}$ is fixed by the normalization condition, $u^2=1$. One can in fact perform the explicit calculations for $u_{\mu}\delta J^{\mu}_{V/5}$ and $u_{\mu}\delta T^{\mu\nu}$ from $\delta f_q$ and show that these terms indeed vanish with hydrodynamic EOM. Also, the hydrodynamic EOM do not affect $T^{\mu\nu}_A$. 


Finally, by implementing 
Eq.(\ref{cons_J_T}) and Eq.(\ref{TAC}), 
we may write down the conservation laws for canonical EM/AM-tensor densities, 
\begin{eqnarray}\label{divTb_totalD}
\partial_{\mu}\bar{T}^{\mu\nu}=\partial_{\mu}T^{\mu\nu}+\partial_{\mu}T_{A}^{\mu\nu}=F^{\nu\rho}J_{V\rho}-\frac{u_{\rho}\delta T^{\rho\nu}}{\tau_R}+\frac{\hbar}{4}\epsilon^{\mu\nu\alpha\beta}\partial_{\mu}\Big(\frac{u_{\alpha}\delta J_{5\perp\beta}}{\tau_R}\Big),
\end{eqnarray}
and
\begin{eqnarray}\label{divMC_totalD}
\partial_{\lambda}M^{\lambda\mu\nu}_{C}=X^{[\mu}F^{\nu]\rho}J_{V\rho}-\frac{u_{\rho}}{\tau_R}X^{[\mu}\delta T^{\rho\nu]}-\frac{\hbar}{4}\partial_{\lambda}\Big(X^{[\mu}\epsilon^{\nu]\lambda\alpha\beta}\frac{u_{\alpha}\delta J_{5\perp\beta}}{\tau_R}\Big).
\end{eqnarray} 
By parameterizing $\delta J_{5\perp\beta}=\tau_R\tilde{J}_{5\perp\beta}$, where $\tilde{J}_{5\perp\beta}$ can be read out from Eq.(\ref{J_V5_noneq}, the above equations can be further written as
\begin{eqnarray}
\partial_{\mu}\bar{T}^{\mu\nu}
=F^{\nu\rho}J_{V\rho}-\frac{u_{\rho}\delta T^{\rho\nu}}{\tau_R}
+\frac{\hbar}{4}\epsilon^{\mu\nu\alpha\beta}u_{\alpha}(\partial_{\mu}-u\cdot\partial u_{\mu})\tilde{J}_{5\perp\beta}+\frac{\hbar}{2}u^{\nu}(\omega\cdot\tilde{J}_{5\perp})
,
\end{eqnarray}
and
\begin{eqnarray}\label{divMC}\nonumber
\partial_{\lambda}M^{\lambda\mu\nu}_{C}
&=&X^{[\mu}F^{\nu]\rho}J_{V\rho}-\frac{u_{\rho}}{\tau_R}X^{[\mu}\delta T^{\rho\nu]}
\\
&&+\frac{\hbar}{2}\Bigg[\epsilon^{\mu\nu\alpha\beta}u_{\alpha}
+X^{[\mu}u^{\nu]}\omega^{\beta}+\frac{u_{\lambda}}{2}X^{[\mu}\epsilon^{\nu]\lambda\alpha\beta}\big(\partial_{\alpha}-u\cdot\partial u_{\alpha}\big)
\Bigg]\tilde{J}_{5\perp\beta}.
\end{eqnarray}
As discussed previously, in relativistic hydrodynamics, the symmetric EM tensor $T^{\mu\nu}$ is required to be conserved except for the coupling between the field strength and the vector-charge currents such that $u_{\rho}\delta T^{\mu\rho}=0$. According to Eq.(\ref{divMC}), it is expected that the electric field can break the AM-momentum conservation. Nonetheless, even when $F^{\mu\nu}=0$, the last term in Eq.(\ref{divMC}) stemming from the non-equilibrium axial-charge current triggered by temperature/chemical-potential gradients still causes a nonzero torque, which locally breaks conservation of the canonical angular momentum. Because the spin current is characterized by an axial-charge current, such a term could be also regraded as a nontrivial spin-orbit interaction. Such an effect also stems from side jumps. Since such a local torque is internal, it should vanish globally when integrating over the position space, which could be seen from Eq.(\ref{divMC_totalD})\footnote{It is clear that the last term of (\ref{divMC_totalD}) will be a surface term when $\lambda=i$ as one of spatial components, which vanishes when integrating over position space. When $\lambda=0$ as the temporal component, the situation is more subtle. Since $\partial_0u_i$ as acceleration of the fluid velocity will be proportional to $\partial_iT$ or $\partial_i\mu$ as the gradients of either temperature or chemical potentials based on the hydrodynamic EOM when $E^{\mu}=0$, the last term in (\ref{divMC_totalD}) should be accordingly proportional to $\epsilon^{ijk}\partial_iT\partial_j\mu$ as the cross product of the gradients of temperature and of chemical potentials. Such a term should also vanish when integrating over position space. The same argument could be applied to (\ref{divTb_totalD}). The net torque should only comes from the external fields.}
Furthermore, in a steady state such that the non-equilibrium vector/axial-charge currents vanish, $\bar{T}^{\mu\nu}$ and $T^{\mu\nu}$ follow the same conservation laws and so do $M^{\lambda\mu\nu}_C$ and $M^{\lambda\mu\nu}_B$. As opposed to $M^{\lambda\mu\nu}_C$, $M^{\lambda\mu\nu}_B$ is locally conserved in the absence of electric fields, which seems to be a better conserved quantity for hydrodynamics, whereas the local angular-momentum transfer through the spin-orbit interaction is not manifested. From Eq.(\ref{divMC_totalD}), one may alternatively define a locally conserved AM tensor in the absence of background fields,
\begin{eqnarray}
\tilde{M}^{\lambda\mu\nu}_C\equiv M^{\lambda\mu\nu}_C+\frac{\hbar}{4}\Big(X^{[\mu}\epsilon^{\nu]\lambda\alpha\beta}\frac{u_{\alpha}\delta J_{5\perp\beta}}{\tau_R}\Big),
\end{eqnarray}
which can be decomposed into the canonical AM tensor and the spin-orbit coupling. The $\tilde{M}^{\lambda\mu\nu}_C$ and $M^{\lambda\mu\nu}_B$ are then connected by the pseudo-gauge transformation \footnote{From the field-theory construction, $M^{\lambda\mu\nu}_C$ and $M^{\lambda\mu\nu}_B$ are related by the pseudo-gauge transformation with the equations of motion. However, in the Wigner-function approach, the equations of motion are Kaddanof-Baym(KB)-like equations instead of the simple Dirac equations in the presence of collisions.}.

It is worthwhile to note that the side jumps in (local) equilibrium do not yield entropy production, which manifests the non-dissipation of the CVE. By carrying out a direct calculation of the entropy-density current from the Wigner function in equilibrium, as shown in Appendix.B, one finds
\begin{eqnarray}
s^{\mu}_{\text{leq}}=\frac{1}{T}\Big(u^{\mu}p+T_{\text{leq}}^{\mu\nu}u_{\nu}-\mu_V J_{V\text{leq}}^{\mu}+\hbar D_{B}B^{\mu}+\hbar D_{\omega}\omega^{\mu}\Big),
\end{eqnarray}
where $D_{B/\omega}=D_{B/\omega R}-D_{B/\omega L}$ and 
\begin{eqnarray}
D_{BR/L}=\frac{1}{8\pi^2}\Big(\mu_{R/L}^2+\frac{\pi^2T^2}{3}\Big)=\frac{\xi_{B R/L}}{T}
,\quad D_{\omega R/L}=\frac{1}{12}\Big(T^2\mu_{R/L}+\frac{\mu_{R/L}^3}{\pi^2}\Big)=\frac{\xi_{\omega R/L}}{2T}.
\end{eqnarray}
The result takes the same form as what has been proposed in anomalous hydrodynamics \cite{Son:2009tf}, in which only the symmetric EM tensor contributes. One can explicitly show that $\partial_{\mu}s^{\mu}_{\text{leq}}=0$. Near local equilibrium, the non-equilibrium fluctuations such as viscous effects will modify $s^{\mu}$ and cause entropy production. It is not clear whether $T^{\mu\nu}_A$ could appear in $s^{\mu}$ for non-equilibrium cases, while such corrections should be at least at $\mathcal{O}(\hbar\partial^2)$ and pertinent to collisions, which might be associated with for example the viscous corrections upon CME/CVE \cite{Hidaka:2018ekt}.

\section{Concluding Remarks and Outlook}\label{conclude_section}
In this paper, we have investigated the interplay between the spin and orbital components of the canonical angular momentum for chiral fluids in the framework of Wigner functions and CKT. It is found that the side jumps result in non-vanishing antisymmetric component of the canonical EM tensor in global equilibrium with nonzero vorticity and an axial chemical potential, which is responsible for the angular-momentum transfer between the spin and fluid. Near local equilibrium, we further obtained the anti-symmetric component depending on temperature/chemical-potentials gradients and electric fields. As indicated by KB-like equations, such contributions are implicitly associated with collisions stemming from the spin-orbit interaction, which further breaks local AM conservation. Also, we have explicitly shown that the the entropy-density current is not affected by the spin-orbit interaction in equilibrium. It thus takes the same form as proposed from anomalous hydrodynamics and causes no entropy production in equilibrium.

In general, we have shown that there exists nontrivial angular-momentum transfer from the spin-orbit interaction up to $\mathcal{O}(\hbar)$ in chiral fluids. However, in the zero axial-charge chemical potential, it is crucial to investigate the similar scenario up to $\mathcal{O}(\hbar^2)$, which incorporates the polarization led by CSE and CVE. The study thus requires future exploration upon the higher-order quantum corrections on Wigner functions and CKT. On the other hand, it is also intriguing to further investigate the polarization spectrum characterized by the Pauli-Lubanski pseudo vector with the quantum corrections from the orbital angular momentum.               

\acknowledgments
This work is supported by the RIKEN Foreign Postdoctoral Researcher program. The author acknowledges K. Fukushima, M. Hongo, S.Pu, and Z. Qiu for fruitful discussions and F. Becattini for reading the manuscript and offering valuable comments. Particularly, the author thanks Y. Hiddaka for providing critical comments and useful suggestions.    

\appendix
\section{Derivation of the Anti-Symmetric EM tensor}
In this Appendix, we present some details of computations for the results shown in the context. Here we only consider right-handed fermions, for which we will omit the subindices $R$ for convenience. 
Based on the Dyson-Schwinger equations under the Wigner transformation up to $\mathcal{O}(\hbar)$, we shall obtain the following Kaddanof-Baym(KB)-like equations for right-handed fermions \cite{Hidaka:2016yjf},
\begin{eqnarray}\label{KB_like}
&&\sigma^{\mu}\left(q_{\mu}+\frac{i\hbar}{2}\Delta_{\mu}\right)\grave{S}^{<}=\frac{i\hbar}{2}\left(\Sigma^<\grave{S}^{>}-\Sigma^>\grave{S}^{<}\right),
\\
&&\left(q_{\mu}-\frac{i\hbar}{2}\Delta_{\mu}\right)\grave{S}^{<}\sigma^{\mu}=-\frac{i\hbar}{2}\left(\grave{S}^{>}\Sigma^<-\grave{S}^{<}\Sigma^>\right),
\end{eqnarray}
By parameterizing $\grave{S}^<=\bar{\sigma}^{\mu}\grave{S}^<_{\mu}$, the above equation yield the difference equations,
\begin{eqnarray}\label{diff_Dirac}\nonumber
&&
\hbar\{\sigma^{\mu},\bar{\sigma}^{\nu}\}\mathcal{D}_{\mu}\grave{S}^<_{\nu}=2i[\sigma^{\mu},\bar{\sigma}^{\nu}]q_{\mu}\grave{S}^<_{\nu},
\\
&&
\hbar[\sigma^{\mu},\bar{\sigma}^{\nu}]\mathcal{D}_{\mu}\grave{S}^<_{\nu}=2i\{\sigma^{\mu},\bar{\sigma}^{\nu}\}q_{\mu}\grave{S}^<_{\nu}.
\end{eqnarray}
where $[A,B]=AB-BA$ and $\{A,B\}=AB+BA$ and
\begin{eqnarray}
\mathcal{D}_{\mu}\grave{S}^{<}_{\nu}=\Delta_{\mu}\grave{S}^{<}_{\nu}-\Sigma^<_{\mu}\grave{S}^{>}_{\nu}+\Sigma^>_{\mu}\grave{S}^{<}_{\nu}
\end{eqnarray}
with $\Delta_{\mu}=\partial_{\mu}+F_{\nu\mu}\frac{\partial}{\partial q_{\nu}}$.
In Eq.(\ref{diff_Dirac}), the traceless part linear to the Pauli matrices yields
\begin{eqnarray}\nonumber\label{traceless_part}
&&\hbar\sigma^{\mu}_{\perp}n^{\nu}\big(\mathcal{D}_{\mu}\grave{S}^<_{\nu}-\mathcal{D}_{\nu}\grave{S}^<_{\mu}\big)=-2\sigma^{\mu}_{\perp}\epsilon_{\alpha\mu\nu\beta}n^{\alpha} q^{\nu}\grave{S}^{<\beta},
\\
&&\hbar\sigma^{\mu}_{\perp}\epsilon_{\alpha\mu\nu\beta}n^{\alpha}\mathcal{D}^{\beta}\grave{S}^{<\nu}=2\sigma^{\mu}_{\perp}\left(q\cdot n\grave{S}^<_{\mu}-q_{\mu}n\cdot\grave{S}^<\right),
\end{eqnarray}
where $V^{\mu}_{\perp}=(\eta^{\mu\nu}-n^{\mu}n^{\nu})V_{\nu}$ for an arbitrary vector $V^{\mu}$ and we set $n\cdot\sigma=I$.
By integrating over momentum space, Eq.(\ref{traceless_part}) becomes
\begin{eqnarray}\nonumber
&&
\frac{\hbar}{2}\Big(n^{\nu}\partial_{\perp\mu}J_{\nu}-n\cdot\partial J_{\perp\mu}-2\int_q\big((q\cdot n)\mathcal{C}_{\perp\mu}-n\cdot \mathcal{C}q_{\perp\mu}\big)\Big)
=\epsilon_{\alpha\mu\beta\nu}n^{\alpha}T_A^{\beta\nu},
\\
&&\frac{\hbar}{2}\epsilon^{\lambda\mu\nu\rho}n_{\nu}\big(\partial_{\lambda}J_{\rho}-2\int_q\mathcal{C}_{\lambda}q_{\rho}\big)=2T^{\mu\nu}_An_{\nu},
\end{eqnarray}
which then results in 
\begin{eqnarray}\label{TA_Collisions}
T^{\mu\nu}_{A}=\frac{\hbar}{4}\epsilon^{\mu\nu\alpha\beta}\Big(\partial_{\alpha}J_{\beta}+2n_{\alpha}\int_q\big((q\cdot n)\mathcal{C}_{\perp\beta}-(n\cdot \mathcal{C})q_{\perp\beta}\big)\Big).
\end{eqnarray}
For left-handed fermions, the $\mathcal{O}(\hbar)$ terms should flip the sign.

The perturbative solution for Wigner functions solved from Eq.(\ref{KB_like}) is shown in Eq.(\ref{Wigner_f}). Near local equilibrium, it is found \cite{Hidaka:2017auj,Hidaka:2018ekt} 
\begin{eqnarray}
\grave{S}^{<\mu}_{\text{leq}}&=&2\pi\bar{\epsilon}(q\cdot u)\Bigg[\delta(q^2)\left(q^{\mu}+\frac{\hbar}{2}\big(u^{\mu}(q\cdot\omega)-\omega^{\mu}(q\cdot u)\big)\partial_{q\cdot u}-\hbar S^{\mu\nu}_{(u)}\tilde{E}_{\nu}
\partial_{q\cdot u}
\right)
\\
&&
+\frac{\hbar}{2}(B^{\mu}u^{\nu}-B^{\nu}u^{\mu}+\epsilon^{\mu\nu\alpha\beta}E_{\alpha}u_{\beta})\partial_{q\nu}\delta(q^2)
\Bigg]f^{(0)}_q,
\end{eqnarray}
where $f^{(0)}_q=(e^{\beta(q\cdot u-\mu_R)}+1)^{-1}$ and we explicitly write down the electric/magnetic-fields dependence via 
\begin{eqnarray}
\frac{1}{2}\epsilon^{\mu\nu\alpha\beta}F_{\alpha\beta}=B^{\mu}u^{\nu}-B^{\nu}u^{\mu}+\epsilon^{\mu\nu\alpha\beta}E_{\alpha}u_{\beta}.
\end{eqnarray}
Here we also introduce the following notations,
\begin{eqnarray}\label{E_t}
\tilde{E}_{\beta}=\mathcal{E}_{\beta}+\frac{(q\cdot u)}{T}\partial_{\beta}T-q^{\sigma}(\sigma_{\beta\sigma}+\kappa_{\beta\sigma}),\quad\mathcal{E}_{\mu}=E_{\mu}+T\partial_{\mu}\bar{\mu},
\end{eqnarray}
where
\begin{eqnarray}
\sigma_{\mu\nu}=(\partial_{\mu}u_{\nu}+\partial_{\nu}u_{\mu})/2,
\quad
\kappa_{\alpha\beta}=\frac{1}{2}\big(u_{\alpha}u\cdot\partial u_{\beta}-u_{\beta}u\cdot\partial u_{\alpha}\big).
\end{eqnarray}
In the local rest frame, $u^{\mu}\approx (1,{\bf 0})$, one finds
\begin{eqnarray}
\grave{S}^{<0}_{\text{leq}}&=&2\pi\bar{\epsilon}(q_0)\Bigg[\delta(q^2)\left(q^{0}+\frac{\hbar}{2}(q\cdot\omega)
\partial_{q0}
\right)
-\frac{\hbar}{2}B^{j}\partial_{qj}\delta(q^2)
\Bigg]f^{(0)}_q,
\end{eqnarray}
and
\begin{eqnarray}\nonumber
\grave{S}^{<i}_{\text{leq}}&=&2\pi\bar{\epsilon}(q_0)\Bigg[\delta(q^2)\left(q^{i}-\frac{\hbar}{2}\omega^{i}q_0\partial_{q0}-\hbar S^{ij}_{(u)}\tilde{E}_{j}
\partial_{q0}
\right)
\\
&&
+\frac{\hbar}{2}(B^{i}\partial_{q0}-\epsilon^{ijk}E_{k}\partial_{qj})\delta(q^2)
\Bigg]f^{(0)}_q.
\end{eqnarray}
We subsequently employ the Wigner functions above to calculate $T^{\mu\nu}_A$. It is found
\begin{eqnarray}\nonumber\label{TA0i}
T_{A\text{leq}}^{0i}&=&-\int \frac{d^4q}{(2\pi)^4}\bar{\epsilon}(q_0)\Big(\grave{S}^{<i}_{R\text{leq}}q^0-\grave{S}^{<0}_{R\text{leq}}q^i\Big)
\\\nonumber
&=&-\frac{\hbar}{2}\int \frac{d^4q}{(2\pi)^3}\bar{\epsilon}(q_0)\Bigg(\delta(q^2)\Big(q^i({\bf q\cdot\bm{\omega}})-q_0^2\omega^i\Big)\partial_{q0}f^{(0)}_q
+\Big(q^0B^i\partial_{q0}\delta(q^2)+q^iB^j\partial_{qj}\delta(q^2)\Big)f^{(0)}_q\Bigg)
\\
&=&-\frac{\hbar\omega^iT^3}{12}\Big(\bar{\mu}_R+\frac{\bar{\mu}_R^3}{\pi^2}\Big)
\end{eqnarray}
and
\begin{eqnarray}\nonumber\label{TAij}
\delta T^{ij}_{A\text{leq}}&=&\int \frac{d^4q}{(2\pi)^4}\bar{\epsilon}(q_0)\Big(\grave{S}^{<i}_{R\text{leq}}q^j-\grave{S}^{<j}_{R\text{leq}}q^i\Big)
\\\nonumber
&=&\hbar\int\frac{d^4q}{(2\pi)^3}\bar{\epsilon}(q\cdot u)\Bigg[\frac{\delta(q^2)}{2q_0}
\big(\epsilon^{iln}q^j-\epsilon^{jln}q^i\big)q_n\tilde{E}_l\partial_{q0}f^{(0)}_q
-q_lE_k\big(\epsilon^{ilk}q^j-\epsilon^{jlk}q^i\big)\frac{\partial\delta(q^2)}{\partial q^2}
f^{(0)}_q\Bigg]
\\
&=&-\delta^{\mu i}\delta^{\nu j}\frac{\hbar\epsilon^{\mu\nu\alpha\beta} u_{\alpha}}{2}\Big(\sigma_{\omega}T\partial_{\beta}\bar{\mu}+N_0\frac{\partial_{\beta}T}{T}-\frac{\sigma_{\omega}\mathcal{E}_{\beta}}{3}\Big).
\end{eqnarray}

\section{The Entropy-Density Current}
We may perform the direct calculation of the entropy-density current through the Wigner function. By introducing the Boltzmann's $\mathcal{H}$ function for fermions (see e.g. Ref.\cite{Loganayagam:2012pz}),
\begin{eqnarray}
\mathcal{H}(f)=-f\ln f-(1-f)\ln(1-f),
\end{eqnarray}  
we may construct the entropy-density current via the Wigner function by replacing $f$ with $\mathcal{H}$,
\begin{eqnarray}
s_R^{\mu}&=&2\int\frac{d^4q}{(2\pi)^3}\bar{\epsilon}(q\cdot n)\Bigg(\delta(q^2)\Big(q^{\mu}
+\hbar S^{\mu\nu}_{(n)}\mathcal{D}_{\nu}\Big)
+\hbar\epsilon^{\mu\nu\alpha\beta}q_{\nu}F_{\alpha\beta}\frac{\partial\delta(q^2)}{2\partial q^2}
\Bigg)\mathcal{H}(f^{(n)}_q),
\end{eqnarray}
where we focus on right-handed fermions. Since we are particularly interested in an equilibrium case, we then take $n^{\mu}=u^{\mu}$ and $f^{(n)}_q=f^{\text{leq}(u)}_{q}=(e^{g}+1)^{-1}$ with $g=\beta(q\cdot u-\mu_R+\hbar q\cdot\omega/(2q\cdot u))$. The entropy current in equilibrium hence becomes
\begin{eqnarray}
s_{R\text{leq}}^{\mu}&=&2\int\frac{d^4q}{(2\pi)^3}\bar{\epsilon}(q\cdot u)\Bigg(\delta(q^2)\Big(q^{\mu}
+\hbar S^{\mu\nu}_{(u)}\Delta_{\nu}\Big)
+\hbar\epsilon^{\mu\nu\alpha\beta}q_{\nu}F_{\alpha\beta}\frac{\partial\delta(q^2)}{2\partial q^2}
\Bigg)\mathcal{H}(f^{\text{leq}(u)}_{q}).
\end{eqnarray} 
It is now more convenient to write the $\mathcal{H}$ function as
\begin{eqnarray}\label{H_exp}
\mathcal{H}(f^{\text{leq}(u)}_{q})=gf^{\text{leq}(u)}_{q}-\ln(1-f^{\text{leq}(u)}_{q}).
\end{eqnarray}
We may first compute the contribution from the first component up to $\mathcal{O}(\hbar)$,
\begin{eqnarray}\nonumber
s_{I}^{\mu}&=&2\int\frac{d^4q}{(2\pi)^3}\bar{\epsilon}(q\cdot u)\Bigg(\delta(q^2)\Big(q^{\mu}
+\hbar S^{\mu\nu}_{(u)}\Delta_{\nu}\Big)
+\hbar\epsilon^{\mu\nu\alpha\beta}q_{\nu}F_{\alpha\beta}\frac{\partial\delta(q^2)}{2\partial q^2}
\Bigg)gf^{\text{leq}(u)}_{q}
\\
&=&\beta\bar{T}^{\mu\nu}_{R\text{leq}}u_{\nu}-\bar{\mu}_R J_{R\text{leq}}^{\mu}+\hbar\int\frac{d^4q}{(2\pi)^3}\bar{\epsilon}(q\cdot u)\delta(q^2)\Big(\frac{\beta (q\cdot\omega)}{q\cdot u}q^{\mu}+2S^{\mu\nu}_{(u)}(\Delta_{\nu}g^{(0)})\Big)f^{(0)}_q
,
\end{eqnarray} 
where $g^{(0)}=\beta(q\cdot u-\mu_R)$. By using
\begin{eqnarray}
2S^{\mu\nu}_{(u)}\Delta_{\nu}g^{(0)}=\frac{\beta}{q\cdot u}\Big(\big(u^{\mu}(q\cdot u)(q\cdot\omega)-q^{\mu}(q\cdot\omega)-\omega^{\mu}(q\cdot u)^2\big)
+\epsilon^{\mu\nu\alpha\beta}q_{\alpha}u_{\beta}\tilde{E}_{\nu}
\Big),
\end{eqnarray} 
we obtain \footnote{When including the contribution from anti-particles, one should keep in mind that the normal ordering is implicitly taken and the corresponding divergent term should be dropped.}
\begin{eqnarray}\nonumber
s_I^{\mu}&=&\beta\bar{T}_{R\text{leq}}^{\mu\nu}u_{\nu}-\bar{\mu}_R J_{R\text{leq}}^{\mu}-\hbar\beta\omega^{\mu}\int\frac{d^4q}{(2\pi)^3}\bar{\epsilon}(q\cdot u)\delta(q^2)(q\cdot u)f^{(0)}_{q}
\\
&=&\beta\bar{T}_{R\text{leq}}^{\mu\nu}u_{\nu}-\bar{\mu}_R J_{R\text{leq}}^{\mu}-\frac{\hbar}{2}\beta N_R\omega^{\mu}.
\end{eqnarray}
Here $\tilde{E}_{\nu}$ is defined in (\ref{E_t}) and one can easily check its contribution vanishes in the integral.
However, as shown in (\ref{TAR_leq}), it is found $T_{AR\text{leq}}^{\mu\nu}u_{\nu}=\hbar N_R\omega^{\mu}/2$. We thus find
\begin{eqnarray}\label{s1}
s^{\mu}_I=\beta T_{R\text{leq}}^{\mu\nu}u_{\nu}-\bar{\mu}_R J_{R\text{leq}}^{\mu},
\end{eqnarray}
which incorporates only the  symmetric EM tensor. Subsequently, we should evaluate the contribution from the second term in (\ref{H_exp}) up to $\mathcal{O}(\hbar)$,
\begin{eqnarray}\nonumber
s^{\mu}_{II}&=&-2\int\frac{d^4q}{(2\pi)^3}\bar{\epsilon}(q\cdot u)\Bigg(\delta(q^2)\Big(q^{\mu}
+\hbar S^{\mu\nu}_{(u)}\Delta_{\nu}\Big)
+\hbar\epsilon^{\mu\nu\alpha\beta}q_{\nu}F_{\alpha\beta}\frac{\partial\delta(q^2)}{2\partial q^2}
\Bigg)\ln (1-f^{\text{leq}(u)}_{q})
\\\nonumber
&=&-2\int\frac{d^4q}{(2\pi)^3}\bar{\epsilon}(q\cdot u)
\Bigg(\Big(\delta(q^2)q^{\mu}+\frac{\hbar}{4}\epsilon^{\mu\nu\alpha\beta}F_{\alpha\beta}\frac{\partial\delta(q^2)}{\partial q^{\nu}}\Big)\ln(1-f^{(0)}_q)
\\
&&
-\frac{\hbar\delta(q^2)}{(1-f^{(0)}_q)}\Big(\frac{q^{\mu}(q\cdot\omega)}{2q\cdot u}\partial_{q\cdot u}+S^{\mu\nu}_{(u)}\Delta_{\nu}\Big)f^{(0)}_q
\Bigg).
\end{eqnarray} 
Performing explicit calculations, we obtain 
\begin{eqnarray}\nonumber\label{s2}
s^{\mu}_{II}&=&-2\int\frac{d^4q}{(2\pi)^3}\delta(q^2)\bar{\epsilon}(q\cdot u)\Bigg(q^{\mu}\ln(1-f^{(0)}_q)-\frac{\hbar\beta}{2}
\big(B^{\mu}+\omega^{\mu}(q\cdot u)-u^{\mu}(q\cdot\omega)+2S^{\mu\nu}_{(u)}\tilde{E}_{\nu}\big)f^{(0)}_q
\Bigg)
\\
&=&\beta u^{\mu}p_R+\frac{\hbar\beta}{8\pi^2}\Big(\mu_R^2+\frac{\pi^2T^2}{3}\Big)B^{\mu}
+\frac{\hbar\beta}{12}\Big(T^2\mu_R+\frac{\mu_R^3}{\pi^2}\Big)\omega^{\mu},
\end{eqnarray}
in which we employ the integration by part to acquire the first term in the second equality from the logarithmic term in the integrand. Similar to the case for $s^{\mu}_I$, the $\tilde{E}_{\nu}$ term does not contribute. Combining (\ref{s1}) and (\ref{s2}), we derive the entropy-density current for right-handed fermions in equilibrium,
\begin{eqnarray}
s_{\text{Rleq}}^{\mu}=s^{\mu}_I+s^{\mu}_{II}=\frac{1}{T}\Big(u^{\mu}p_R+T_{R\text{leq}}^{\mu\nu}u_{\nu}-\mu_R J_{R\text{leq}}^{\mu}+\hbar D_{BR}B^{\mu}+\hbar D_{\omega R}\omega^{\mu}\Big),
\end{eqnarray}  
where 
\begin{eqnarray}
D_{BR}=\frac{1}{8\pi^2}\Big(\mu_R^2+\frac{\pi^2T^2}{3}\Big)=\frac{\xi_{BR}}{T},\quad D_{\omega R}=\frac{1}{12}\Big(T^2\mu_R+\frac{\mu_R^3}{\pi^2}\Big)=\frac{\xi_{\omega R}}{2T}.
\end{eqnarray} 
One may further compute $s_{\text{Lleq}}^{\mu}$ for left-handed fermions, where the $\mathcal{O}(\hbar)$ terms flip the signs, and obtain the total entropy-density current $s_{\text{leq}}^{\mu}=s_{\text{Rleq}}^{\mu}+s_{\text{Lleq}}^{\mu}$.

\bibliography{polarization_chiral_fluids_v3.bbl}

\end{document}